\documentclass[twocolumn,pra,aps,floatfix,longbibliography]{revtex4-1}
\usepackage{amsmath,esint}
\usepackage{amsfonts}
\usepackage{amssymb}
\usepackage{gensymb}
\usepackage{siunitx}
\usepackage{verbatim}
\usepackage{graphicx}
\usepackage{xfrac}
\usepackage{times}
\usepackage{natbib}
\usepackage{hyperref}
\usepackage{tabularx}
\usepackage{dsfont}
\usepackage{url}
\usepackage{appendix}
\usepackage{multirow}
\usepackage[table]{xcolor}
\definecolor{LightCyan}{rgb}{0.95,1,1}

\newcolumntype{a}{>{\columncolor{LightCyan}}X}

\begin{document}

\title{Optical $N$-invariant of graphene's viscous Hall fluid}

\author{Todd Van Mechelen}
\affiliation{Purdue University, School of Electrical and Computer Engineering, Brick Nanotechnology Center, 47907, West Lafayette, Indiana, USA}
\author{Zubin Jacob}
\email{zjacob@purdue.edu}
\affiliation{Purdue University, School of Electrical and Computer Engineering, Brick Nanotechnology Center, 47907, West Lafayette, Indiana, USA}

\begin{abstract}


Over the past three decades, graphene has become the prototypical platform for discovering unique phases of topological matter. Both the Chern insulator $C\in\mathbb{Z}$ and the quantum spin Hall insulator $\nu\in\mathbb{Z}_2$ were first predicted in graphene, which led to a veritable explosion of research in topological materials. Here, we introduce a new topological classification of two-dimensional matter -- the optical $N$-phases $N\in\mathbb{Z}$. The $C$ and $\nu$ phases are related to charge and spin transport respectively, whereas the $N$-phases are connected to polarization transport. In all three cases, transportation of charge/spin/polarization quanta is forbidden in the bulk but permitted on the edge. One fundamental difference is that the $N$-invariant is defined for dynamical electromagnetic waves over all Matsubara frequencies and wavevectors. We show this topological quantum number is captured solely by the spatiotemporal dispersion of the susceptibility tensor $\chi(\omega,\mathbf{q})$. We also prove $N\neq 0$ is nontrivial in graphene's viscous Hall fluid with the underlying physical mechanism being Hall viscosity $\eta_H$. In the nontrivial phase, we discover a deep sub-wavelength phenomenon reminiscent of the Meissner effect: at a particularly large photon momentum $q=D_H^{-1}$ defined by the Hall diffusion length $D_H$, the magnetic field is completely expelled from the viscous Hall fluid. We propose a new probe of topological matter, evanescent magneto-optic Kerr effect (e-MOKE) spectroscopy, to unravel this novel optical $N$-invariant and verify the magnetic field expulsion. Our work indicates that graphene with Hall viscosity is the first candidate material for a topological electromagnetic phase of matter.

\end{abstract}

\maketitle 

\section{Introduction}

Monolayer graphene forms the canonical system to study two-dimensional topological phases of matter. The now famous Haldane model of graphene \cite{Haldane1988}, with time-reversal breaking next-nearest-neighbor (NNN) hopping, was the first proposal of a Chern phase $C\in\mathbb{Z}$. Conversely, the Kane-Mele model \cite{Kane2005} preserves time-reversal symmetry and was the first example of a quantum spin Hall phase $\nu\in\mathbb{Z}_2$, resulting from spin-orbit coupling in graphene. Nevertheless, the growing collection of topological phases in condensed matter can be categorized under the umbrella of electrostatics since all observables, e.g. the quantum Hall $\sigma_{xy}=Ce^2/h$ and spin Hall $\sigma_{xy}^s=\nu e/2\pi$ conductivity, are interpreted at zero photon energy and momentum $\omega=q=0$. One must go beyond this paradigm to characterize the optical properties of matter, as these are defined for electromagnetic fluctuations over all frequencies $\omega\neq 0$ and momenta $q\neq 0$. We are quickly confronted with two important questions: what are the optical invariants of a material? Do these topological invariants represent unique electromagnetic phases of matter? Our work lays the foundations for this optical classification of condensed matter.

\begin{figure*}
\includegraphics[width=0.8\linewidth]{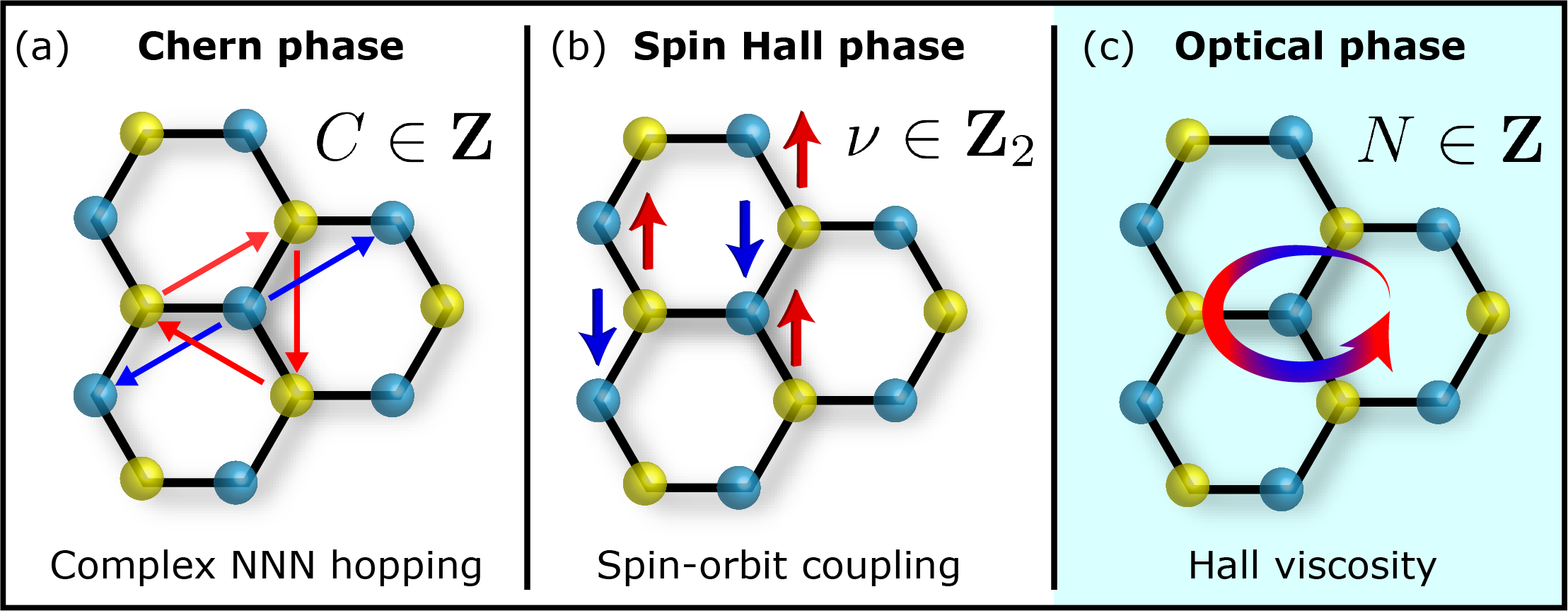}
\caption{Topological phases of graphene. (a) The Chern phase $C\in\mathbb{Z}$ arises from complex next-nearest-neighbor (NNN) hopping. (b) The quantum spin Hall phase $\nu\in\mathbb{Z}_2$ is due to spin-orbit coupling (topological insulator). (c) The optical phase $N\in\mathbb{Z}$ we put forth in this paper is a consequence of Hall viscosity. These three phases can be identified as the Chern insulator, quantum spin Hall insulator and viscous Hall insulator respectively.}
\label{fig:graphene_phases}
\end{figure*}

In this paper, we present graphene imbued with Hall viscosity $\eta_H$ as a paradigmatic example of a topological electromagnetic phase of matter. Please see Fig.~\ref{fig:graphene_phases} comparing the fundamental differences between the Chern phase, quantum spin Hall phase and the optical $N$-phase. This nontrivial topology is revealed in the magnetohydrodynamics of the two-dimensional Navier-Stokes equations. Until quite recently however, viscous electrohydrodynamics with nonzero magnetic field $B\neq 0$ has been experimentally inaccessible, mainly due to impurities and electron-phonon scattering \cite{Berdyugin162}. Exceptional grade 2D materials like graphene \cite{Lucas_2018,Ho2018,Crossno1058} are providing the first platforms to study the fluidic behavior of electrons \cite{Muller2009,Bandurin1055,Bandurin2018}. These viscous Hall fluids have been a theoretical object for some time due to the intriguing phenomenon known as Hall viscosity $\eta_H$, the dissipationless component of the viscous stress tensor \cite{Mendoza2013,Moore2017,Bradlyn2012,Soni2019}. Although difficult to measure, Hall viscosity is a generic feature of parity and time-reversal breaking fluids and can exhibit quantization analogous to the Hall conductivity \cite{Avrom1995,Hoyos2012,Sherafati2016}. Multiple possible observables have been proposed to identify Hall viscosity experimentally, such as negative nonlocal resistance \cite{Levitov2016,Alekseev2016} and anomalous Hall resistivity \cite{Delacretaz2017,Pellegrino2017}. Nevertheless, most studies on Hall viscosity have focused on the steady state (electrostatic) properties of the material; the optical response of the viscous Hall fluid has remained almost completely unexplored \cite{QuantGyro2018,Nonlocal2019,Todd2019}. We show that the Hall viscosity $\eta_H$ actually defines a distinct topological phase of matter $N\in\mathbb{Z}$, which only manifests in the optical regime $\omega\neq 0$.

An optical phase $N\in\mathbb{Z}$ is characterized by the winding number of the susceptibility tensor $\chi(\omega,\mathbf{q})$ -- or equivalently the conductivity tensor $\sigma(\omega,\mathbf{q})$. This topological invariant is found by integrating the material response over all Matsubara frequencies and wavevectors of the photon. We emphasize that the optical $N$-invariant only has meaning for spatiotemporally dispersive media; we must consider nonlocal effects from the outset. Using elementary physical constraints, we prove that $N$ is generically quantized and immune to perturbations in the optical response $\chi\to\chi+\delta\chi$. Our definition utilizes a Volovik (Green's function) formalism \cite{volovik2009universe,Volovik_2017,Gurarie2011,Essin2011} which is naturally generalized to quantum, dissipative and finite temperature systems \cite{Hirsbrunner2019,Kawabata2019}. Although we only consider the continuum theory here, the formalism is robust and easily extended to the lattice case. We argue that $N$ is the central topological quantity of the electromagnetic linear response theory and classifies all two-dimensional optical media with broken time-reversal symmetry. It bears repeating that a nontrivial Chern invariant $C\neq 0$ \textit{does not} imply a nontrivial optical invariant $N\neq 0$, and vice versa [Tbl.~\ref{tab:grapehen_phases}].

We also discover a unique magnetic field repulsion that occurs for deep sub-wavelength photons interacting through Hall viscosity. The complete expulsion of the magnetic field is reminiscent of the Meissner effect and is the smoking gun of a topological electromagnetic phase of matter. To probe this sub-wavelength phenomenon and novel topological signature of matter, we propose evanescent magneto-optic Kerr effect (e-MOKE) spectroscopy. The e-MOKE angle $\Theta$, that measures the rotation of the reflected polarization from the viscous Hall fluid, is a direct observable of the gyrotropic response. If the e-MOKE angle $\Theta(q)=0$ passes through zero at a critical photon momentum $q=D_H^{-1}$, the viscous Hall fluid is topologically nontrivial $N\neq 0$. We show that this is the momentum where the magnetic field is completely expelled from the fluid and only occurs in the nontrivial regime. Here, $D_H=\sqrt{\nu_H/\omega_c}$ defines the Hall diffusion length. 

Our work merges the fields of topological photonics and condensed matter physics to spawn a new area of research in materials science. Although topological photonics \cite{Lu2016,Stone_2016,Horsley2019,Pakniyat2020} has mainly focused on artificial media like photonic crystals \cite{Yuan2018,Wang2019,Cerjan2019} and metamaterials \cite{Liu2019,Peng2019}, our findings demonstrate that natural media can also host topological electromagnetic states. As such, the optical phases we discuss here are properties of matter itself and are not related to some form of macroscopic engineering. The optical invariant $N\in\mathbb{Z}$ is therefore a classification of different topological phases of matter.  

\textbf{Note:} To avoid confusion, we adopt two vector conventions throughout the paper. We denote three-dimensional (3D) vectors with arrows $\Vec{\mathcal{A}}=(\mathcal{A}_x,\mathcal{A}_y,\mathcal{A}_z)$ and reserve boldface for two-dimensional (2D) vectors $\mathbf{A}=(A_x,A_y)$ which corresponds to strictly in-plane components.

\begin{table}
\caption{Summary of the 2+1D topological phases in graphene. The charge $C$ and spin $\nu$ phases are defined at zero photon energy and momentum $\omega=q=0$. The optical phases $N$ are defined for dynamical electromagnetic (EM) fields, $\omega\neq 0$ and $q\neq0$.}
\label{tab:grapehen_phases}
\begin{tabularx}{\linewidth}{X|XXa}
\hline\hline
Quanta & (a) Charge & (b) Spin & (c) Polarization\\ \hline
Class & A & AII &  D \\
Invariant &$C\in\mathbb{Z}$ & $\nu\in\mathbb{Z}_2$ &  $N\in\mathbb{Z}$ \\
Mechanism & NNN hopping & Spin-orbit coupling &  Hall viscosity \\
Observable&$\sigma_{xy}=Ce^2/h$ & $\sigma_{xy}^s=\nu e/2\pi$ &  $B$-field repulsion\\\hline
EM field & $\omega=q=0$ & {}  &  $\omega\neq 0$,  $q\neq0$\\
\hline\hline
\end{tabularx}
\end{table}

\section{Viscous Hall Fluid}\label{sec:LinearHydro}

Our starting point is the two-dimensional Navier-Stokes (NS) equations subject to a uniform magnetic field $B$ and a spatiotemporally varying electric field $\mathbf{E}$. In  Fermi-liquid theory, the 2D NS equations describe the viscous flow of a parity and time-reversal breaking Hall fluid \cite{Banerjee2017,Souslov2019}. This theory has successfully explained the steady state viscous properties of graphene \cite{Berdyugin162}. On the other hand, our analysis focuses on the dynamical time-dependent behavior. Viscosity characterizes the resistance to deformation and amounts to a restoring force in the NS equations. The conventional shear viscosity $\eta$ is dissipative (frictional) and impedes the motion of the fluid. The Hall viscosity $\eta_H$ however, is dissipationless and generates a force perpendicular to the motion. Assuming the electric field fluctuations are relatively weak, the charge density $\varrho=\varrho_0+\delta\varrho$ will be perturbed around its equilibrium value $\varrho_0=-en_0$, where $e$ is the elementary charge and $n_0$ is the electron density. We derive the linearized time-dependent NS equations \cite{castellanos_electrohydrodynamics_1998}, which incorporates acceleration $\partial_t\mathbf{J}\neq\mathbf{0}$ and compressibility $\pmb{\nabla}\cdot\mathbf{J}\neq 0$ of the Hall fluid,
\begin{widetext}
\begin{equation}\label{eq:Hyrdoynamic}
\partial_t \mathbf{J}=-v^2_s\pmb{\nabla}\varrho-(\gamma-\nu\pmb{\nabla}^2)\mathbf{J}-(\omega_c+\nu_H\pmb{\nabla}^2)\mathbf{J}\times\hat{z}+c\frac{\omega_p}{4\pi}
\mathbf{E}.
\end{equation}
\end{widetext}
The complete derivation is outlined in the supplementary information. $\mathbf{E}(t,\pmb{\rho})=\Vec{\mathcal{E}}_\parallel(t,\pmb{\rho},0)$ represents the parallel electric field at the location of the electron fluid $z=0$, where $\pmb{\rho}=(x,y)$ are the in-plane coordinates. Combining Eq.~(\ref{eq:Hyrdoynamic}) with the continuity equation $\partial_t\varrho+\pmb{\nabla}\cdot\mathbf{J}=0$ completely specifies the charge $\varrho$ and current $\mathbf{J}$ densities with appropriate boundary conditions. $v_s\simeq v_F/\sqrt{2}$ is the speed of sound \cite{giuliani2005quantum} which is proportional to the Fermi velocity $v_F$ and $\gamma=\tau^{-1}$ is the phenomenological damping rate. We have assumed linear dispersion characteristic of graphene \cite{FETTER1973367,Kolomeisky2017} to obtain the proportionality of $v_s$ to the Fermi velocity $v_F$. $\omega_p= 4\pi e^2n_0/(mc)$ is the plasma frequency of a 2D electron fluid and $\omega_c=eB/(mc)$ is the cyclotron frequency. Here, $m$ is the effective mass of the electron and $c$ is the speed of light. $\nu=\eta/(m n_0)$ and $\nu_H=\eta_H/(m n_0)$ are the kinetic shear and kinetic Hall viscosities respectively. We can also define two important length scales: the shear $D_\nu=\sqrt{\nu\tau}$ and Hall $D_H=\sqrt{\nu_H/\omega_c}$ diffusion lengths which characterize the hydrodynamic behavior at mesoscopic scales. 

The relative sign of $\nu_H$ with respect to the cyclotron frequency $\omega_c$ is paramount to the topological physics and dictates whether the Hall viscosity repels $\omega_c\nu_H>0$ or reinforces $\omega_c\nu_H<0$ the magnetic field. In a semiclassical electron fluid, the kinetic viscosities $\nu$ and $\nu_H$ are related \cite{Pellegrino2017,Moore2017},
\begin{equation}
\nu=\frac{B_0^2}{B^2+B_0^2}\nu_0, \qquad \nu_H=\frac{BB_0}{B^2+B_0^2}\nu_0,
\end{equation}
where $\nu_0\geq 0$ is the kinematic shear viscosity at zero bias $B=0$, and $B_0=cn_0/(4\pi e\mathcal{N}_0\nu_0)$ is an intrinsic magnetic field of the electron fluid. $\mathcal{N}_0$ being the density of states at the Fermi energy. Notice that the product of $\omega_c$ and $\nu_H$ is positive definite $\omega_c\nu_H>0$; the Hall viscosity repels the magnetic field for all values of $B$. The Hall diffusion length $D_H$ is therefore a real mesoscopic scale of the system. 

\subsection{Magnetohydrodynamic susceptibility}
\label{sec:LinearResponse}

We now derive the bulk linear response theory of an unbounded viscous Hall fluid. Assuming translational symmetry in the $\pmb{\rho}=(x,y)$ plane, the in-plane momentum $\mathbf{q}=(q_x,q_y)$ is conserved which means we can Fourier transform to the reciprocal space. Due to nonlocality arising from pressure $v_s\neq 0$ and viscosity $\nu\neq 0$, the momentum space is particularly useful to understand the linear response theory. To facilitate this, we utilize the susceptibility tensor $\chi$,
\begin{equation}
\mathbf{P}(\omega,\mathbf{q})=\chi(\omega,\mathbf{q})\cdot\mathbf{E}(\omega,\mathbf{q}),
\end{equation}
which gives the induced polarization density $\mathbf{P}$ to an applied electric field $\mathbf{E}$. The response function $\chi$ completely characterizes the bulk optical properties of the material, for every energy $\omega$ and momentum $\mathbf{q}$ of the photon. Note that both $\omega$ and $\mathbf{q}$ are real parameters here. Exploiting rotational symmetry, we derive the components of the susceptibility tensor in an orthogonal basis,
\begin{equation}
\chi_{ij}=\chi_t(\delta_{ij}-\hat{q}_i\hat{q}_j)+\chi_l\hat{q}_i\hat{q}_j+ig\epsilon_{ij},
\end{equation}
where $\hat{q}_i=q_i/q$ is the unit vector directed along the in-plane momentum and $q=\sqrt{\mathbf{q}\cdot\mathbf{q}}$ is its magnitude. $\delta_{ij}$ is the identity and $\epsilon_{ij}$ is the Levi-Civita symbol. We can also translate to the conductivity tensor $\chi=i\sigma/\omega$ if one prefers to work with the induced current density $\mathbf{J}$.

The response function is temporally $\omega$ and spatially $\mathbf{q}$ dispersive, and both properties are essential to realize optical $N$-phases. Temporal dispersion is necessary to establish the electronic band gap $E_\mathrm{bg}$, while spatial dispersion characterizes the geometric phase of the Hall fluid. We decompose the components of $\chi$ into a transverse $\chi_t$,
\begin{subequations}\label{eq:susept}
\begin{equation}
4\pi \chi_t=-\frac{c\omega_p}{\omega\tilde{\omega}}\left(1+\frac{\omega\Omega_c^2}{\omega(\tilde{\omega}^2-\Omega_c^2)-v^2_sq^2\tilde{\omega}}\right),
\end{equation}
longitudinal $\chi_l$,
\begin{equation}
4\pi \chi_l=-\frac{c\omega_p\tilde{\omega}}{\omega(\tilde{\omega}^2-\Omega_c^2)-v^2_sq^2\tilde{\omega}},
\end{equation}
and gyrotropic $g$ response,
\begin{equation}
4\pi g=\frac{c\omega_p\Omega_c}{\omega(\tilde{\omega}^2-\Omega_c^2)-v^2_sq^2\tilde{\omega}}.
\end{equation}
\end{subequations}
Here, $\tilde{\omega}=\omega+i\Gamma$ is the shifted energy, where $\Gamma(q)=\gamma+\nu q^2$ is the viscous damping rate. $\Omega_c$ is the viscous cyclotron frequency,
\begin{equation}\label{eq:VCF}
\Omega_c(q)=\omega_c-\nu_Hq^2.
\end{equation}
Due to Hall viscosity $\nu_H$, the effective magnetic field in the Hall fluid is momentum dependent $B_\mathrm{eff}(q)=mc\Omega_c(q)/e=B(1-D_H^2q^2)$ and varies on the scale of the Hall diffusion length $D_H=\sqrt{\nu_H/\omega_c}$. In the dissipationless (Hermitian) limit $\Gamma\to0$, we obtain the linear response theory of an ideal quantum Hall fluid $\chi=\chi^\dagger$.

Lastly, we verify that the susceptibility tensor satisfies the reality condition,
\begin{equation}\label{eq:Reality}
\chi^*(-\omega,-\mathbf{q})=\chi(\omega,\mathbf{q}),
\end{equation}
since electromagnetism is a real-valued field theory. Due to Eq.~(\ref{eq:Reality}), the components of $\chi$ cannot be completely independent, implying the excitations belong to universality class D \cite{Ryu_2010}. This should be contrasted with the electron, which belongs to class A (or class AII in the presence of time-reversal symmetry). Generically, the spatiotemporally dispersive susceptibility tensor $\chi(\omega,\mathbf{q})$ represents a mapping from the 2+1D momentum space to the general real linear group $\mathrm{GL}_n(\mathbb{R})$. $n$ denotes the degrees of freedom - i.e. the total number of bulk plasmon/exciton modes (poles and zeros of $\det\chi$).

\section{Optical $N$-invariant}
\label{sec:VRI}

\begin{figure}
\includegraphics[width=0.8\linewidth]{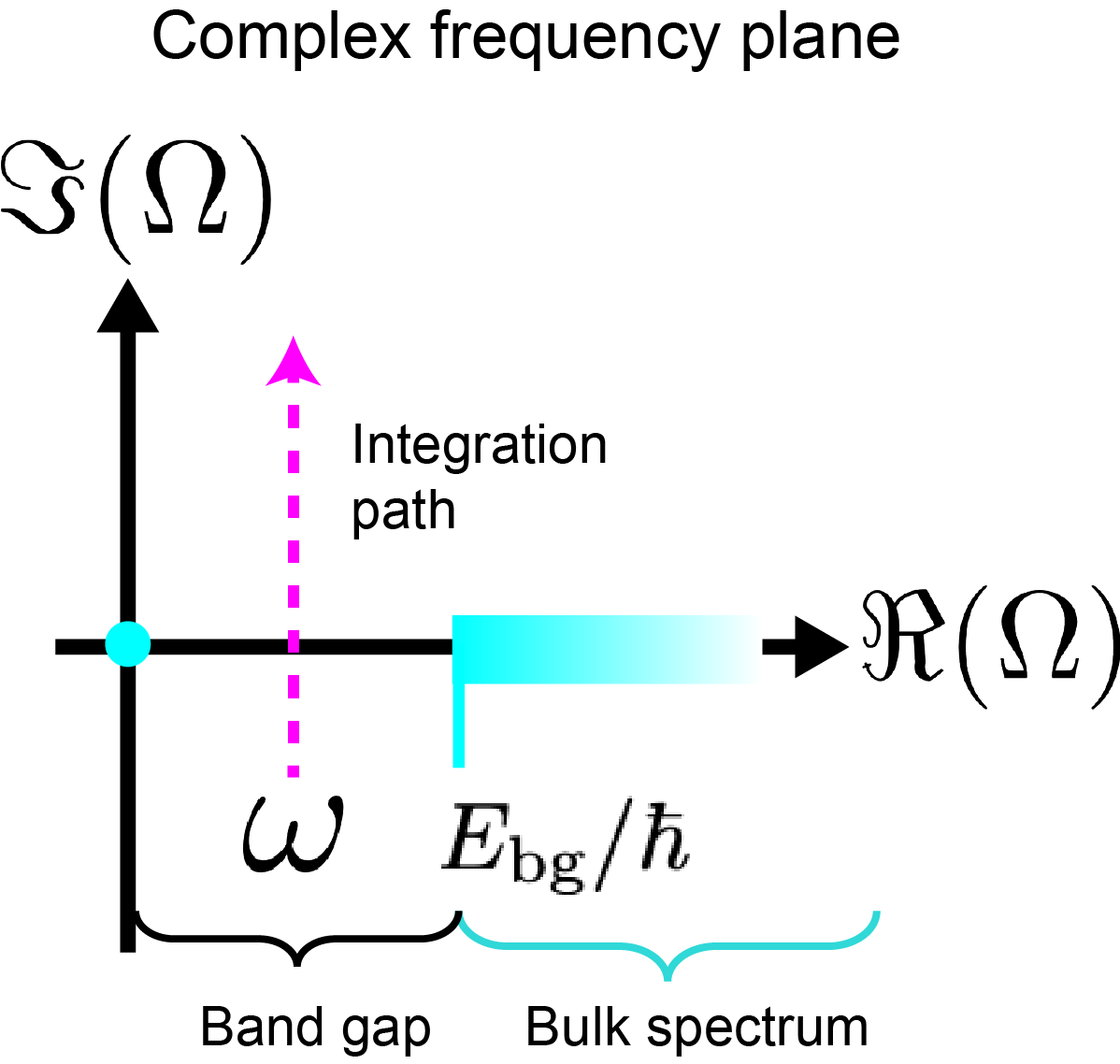}
\caption{The contour used to evaluate the optical $N$-invariant. Integration is performed vertically $\Omega\in (\omega-i\infty,\omega+i\infty)$ over all imaginary (Matsubara) frequencies. $\omega$ is the photon energy that lies within the electronic band gap $0<\hbar\omega<E_\mathrm{bg}$.}
\label{fig:contour}
\end{figure}

The susceptibility tensor $\chi$ is precisely the Green's function of the polarization density $\mathbf{P}$ and is therefore a topological object. The cornerstone of the Green's function approach developed by Volovik \cite{volovik2009universe,Volovik_2017} and Gurarie \cite{Gurarie2011,Essin2011}, lies the following 2+1D topological invariant,
\begin{equation}\label{eq:Volovik}
N=\frac{\epsilon^{\alpha\beta\gamma}}{24\pi^2}\int d^3q~\mathrm{tr}\left[\chi\frac{\partial\chi^{-1}}{\partial q_\alpha}\chi\frac{\partial\chi^{-1}}{\partial q_\beta}\chi\frac{\partial\chi^{-1}}{\partial q_\gamma}\right],
\end{equation}
where $\chi(\omega,\mathbf{q})\to\chi(\Omega,\mathbf{q})$ is parameterized by the complex frequency variable $\omega\to\Omega$. In this case, $q_\alpha=(\Omega,q_x,q_y)$ is the total momentum coordinate and $d^3q=d\Omega d\mathbf{q}$ is the volume element in reciprocal space. $\mathrm{tr}$ denotes the trace over the tensor indices. The temporal integral $d\Omega$ is performed \textit{vertically} over all imaginary (Matsubara) frequencies,
\begin{equation}\label{eq:contour}
\Omega\in(\omega-i\infty,\omega+i\infty).
\end{equation}
$\omega=\Re (\Omega)$ is the photon energy that is assumed to lie within the electronic band gap $0<\hbar\omega< E_\mathrm{bg}$. As such, bulk current cannot be generated since the photon does not possess sufficient energy to excite a plasmon/exciton. With parabolic dispersion $2\omega_c\nu_H<v_s^2$, the band gap of an ideal quantum Hall fluid [Eq.~(\ref{eq:susept})] is defined by the first Landau level $E_\mathrm{bg}=\hbar\Omega_c(0)=\hbar\omega_c$. A visualization of the contour integral is shown in Fig.~\ref{fig:contour}.

We reiterate that the optical $N$-invariant [Eq.~(\ref{eq:Volovik})] is fundamentally different than the electronic Chern number $C$ as they describe physically distinct quantities. The Chern number $C$ is related to the Berry phase of the wave function $\psi$. On the other hand, the $N$-invariant quantifies the geometric phase accumulated from fluctuations in the polarization density $\mathbf{P}$. To prove $N$ is a quantized topological invariant, we consult a few fundamental properties of the response function. All materials are transparent at $|\Omega|\to\infty$ since the electrons cannot respond to rapidly varying temporal oscillations. The response function $\chi$ decays at least as fast as $|\Omega|^{-1}$,
\begin{equation}\label{eq:transparency}
\lim_{|\Omega|\to\infty}\chi(\Omega,\mathbf{q})\to 0.
\end{equation}
By including the point at infinity $|\Omega|=\infty$, each contour in Eq.~(\ref{eq:contour}) defines a circle $S^1$ on the extended complex plane (Riemann sphere). Moreover, due to Hall viscosity $\nu_H\neq0$, the susceptibility tensor is naturally regularized \cite{Ryu_2010} and approaches a directionally independent value as $q\to\infty$,
\begin{equation}\label{eq:regular}
\lim_{q\to \infty}\chi(\Omega,\mathbf{q})\to\chi(\Omega,q).
\end{equation}
Again, by including the point at infinity $q=\infty$, the $\mathbf{q}$ space is equivalent to the sphere $S^2$. The combined 2+1D momentum space $q_\alpha=(\Omega,q_x,q_y)$ is effectively $S^1\times S^2$; i.e. a closed manifold. The spatiotemporally dispersive susceptibility tensor $\chi(\Omega,\mathbf{q})$ is therefore an element of the third homotopy group of $\mathrm{GL}_n(\mathbb{R})$,
\begin{equation}
\pi_3[\mathrm{GL}_n(\mathbb{R})]=\mathbb{Z},
\end{equation}
which is isomorphic to $\mathbb{Z}$ \cite{nakahara2003geometry}. Equation (\ref{eq:Volovik}) calculates the precise element of $N\in \mathbb{Z}$ the response function corresponds to, where each integer represents a unique optical phase. This should be contrasted with the electronic Chern number $C$, which is characterized by complex fields $\pi_3[\mathrm{GL}_n(\mathbb{C})]=\mathbb{Z}$. We emphasize that although both $N$ and $C$ are integer invariants, the symmetry classes are fundamentally different.

To demonstrate explicitly that $N$ is quantized, we consider an arbitrary perturbation to the response function $\chi\to\chi+\delta\chi$ which takes $N\to N+\delta N$. In the dissipationless limit $\Gamma\to 0$, all perturbations amount to a total divergence in the 2+1D momentum such that $\delta N=0$ vanishes identically \cite{Hirsbrunner2019}. $N\in\mathbb{Z}$ is thus topologically quantized and immune to perturbations in the optical response. The full proof is presented in the supplementary information. In a nontrivial regime $N\neq 0$, we expect gapless chiral edge currents to appear. These edge modes are optically excitable within the electronic band gap $0<\hbar\omega<E_\mathrm{bg}$.

\begin{figure}
\includegraphics[width=\linewidth]{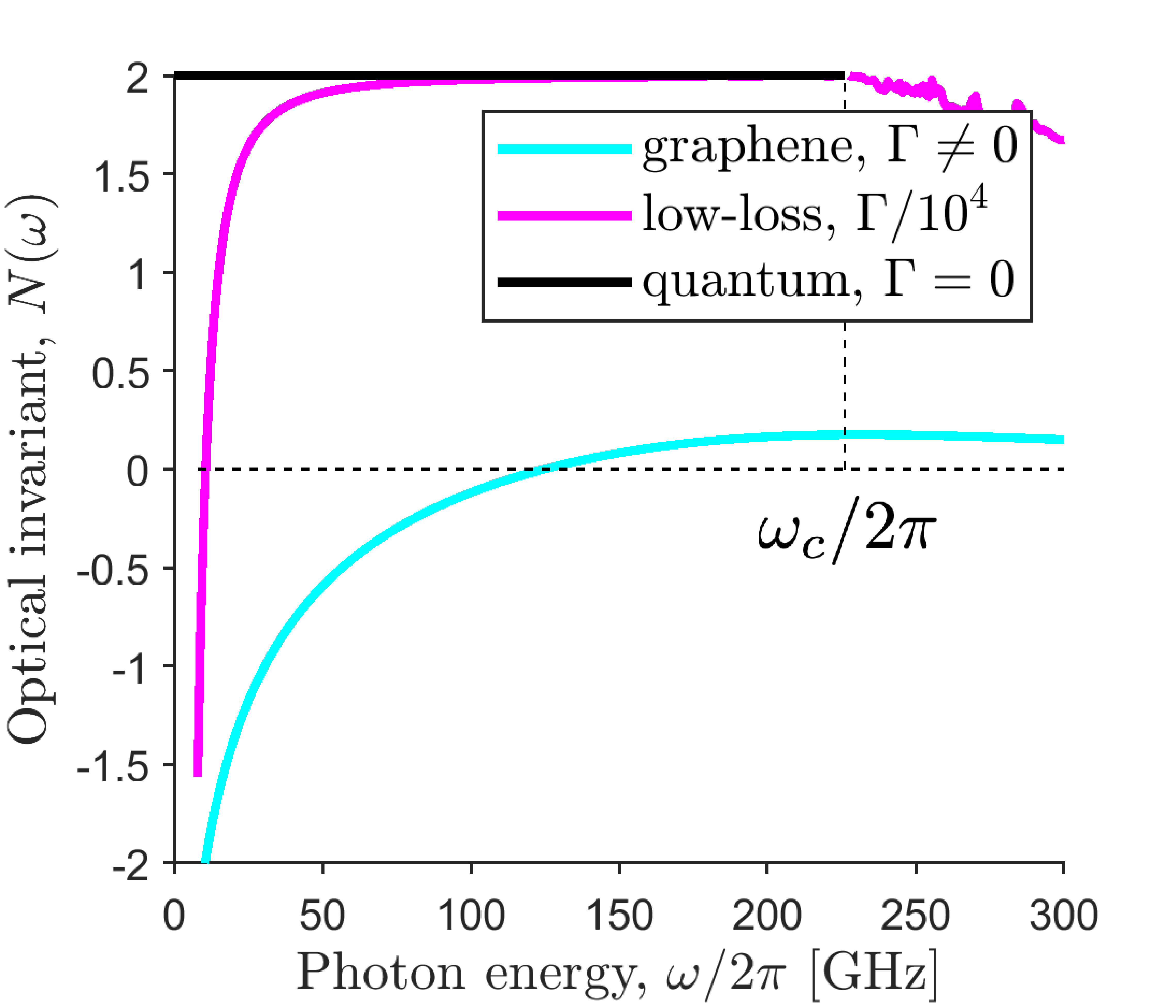}
\caption{Optical invariant $N=N(\omega)$ as a function of the photon energy $\omega$. $N$ is calculated for graphene, low-loss, and quantum Hall fluids for various values of $\omega$. In quantum systems $\Gamma=0$, the optical invariant is quantized to $|N|=2$ in the nontrivial (repulsive) regime $\omega_c\nu_H>0$ when the photon energy $0<\omega<\omega_c$ lies within the electronic band gap. }
\label{fig:Invariant}
\end{figure}

\subsection{Optical $N$-invariant of the viscous Hall fluid}

We now prove that graphene's viscous Hall fluid is the first candidate material for a topological electromagnetic phase of matter. Specifically, it is nontrivial $N\neq 0$ when the Hall viscosity is repulsive $\omega_c\nu_H>0$. Using the magnetohydrodynamic theory [Eq.~(\ref{eq:susept})] and the homotopy equation [Eq.~(\ref{eq:Volovik})] we arrive at,
\begin{equation}\label{eq:ResponseCurvature2}
N=-i\int \frac{d^3q}{2\pi^2} \frac{v^4_sq^2}{\Omega}\frac{[\Omega(\omega_c+\nu_Hq^2)+i(\gamma\nu_H+\omega_c\nu)q^2]}{[\Omega(\tilde{\Omega}^2-\Omega_c^2)- v^2_sq^2\tilde{\Omega}]^2},
\end{equation}
where $\tilde{\Omega}=\Omega+i\Gamma$ and the sign of the dissipation is implied $\Gamma\to\mathrm{sgn}[\Im(\Omega)]\Gamma$ to preserve causality. We integrate Eq.~(\ref{eq:ResponseCurvature2}) numerically in Fig.~\ref{fig:Invariant} but it is important to confirm that $N$ is quantized in the low temperature quantum limit $\Gamma\to0$. Integrating over all Matsubara frequencies and wavevectors, we acquire the optical $N$-invariant of the viscous Hall fluid,
\begin{equation}\label{eq:QuantumVRI}
N=\mathrm{sgn}(\omega_c)+\mathrm{sgn}(\nu_H).
\end{equation}
The invariant is quantized to $|N|=0$ or $2$, when the photon energy $0<\omega<\omega_c$ lies in the electronic band gap. In the nontrivial phase $|N|=2$, we expect chiral edge currents to be optically excitable within the electronic gap $0<\omega<\omega_c$. Nevertheless, since optical properties are directly observable, we can also probe the bulk medium for interesting topological effects.  

\begin{table}
\caption{Parameters for monolayer graphene e-MOKE.}
\label{tab:graphene}
\begin{tabularx}{\linewidth}{Xc}
\hline\hline
Lattice constant, $a$ & $2.46~\si{\angstrom}$ \\
Electron density, $n_0$          & $2\times10^{12}~\mathrm{cm}^{-2}$   \\
Effective electron mass, $m$ & $ 0.0124m_e $ \\
Plasma frequency, $\omega_p/2\pi$     &$2.73~\mathrm{THz}$ \\
Fermi velocity, $v_F$            & $1.1\times10^6~\mathrm{m/s}$                 \\

Kinematic viscosity (zero bias), $\nu_0$    & $0.1~\mathrm{m^2/s}$                \\
Intrinsic magnetic field, $B_0$ & $0.2~\mathrm{T}$                    \\
Biasing magnetic field, $B$     & $0.1~\mathrm{T}$                      \\

Transport time, $\tau$           & $2~\mathrm{ps}$                     \\
Kinetic shear viscosity, $\nu$                & $0.08~\mathrm{m^2/s}$ \\
Shear diffusion length, $D_\nu$  & $0.4~\mu\mathrm{m}$            \\

Cyclotron frequency, $\omega_c/2\pi$  & $226~\mathrm{GHz}$ \\
Kinetic Hall viscosity, $\nu_H$         & $0.04~\mathrm{m^2/s}$ \\
Hall diffusion length, $D_H$         & $0.17~\mu\mathrm{m}$                  \\

Vacuum index, $n_+$  & $1$            \\
Substrate index, $n_-$  & $10$            \\
\hline\hline
\end{tabularx}
\end{table}

\section{\texorpdfstring{\MakeLowercase{e}}{e}-MOKE spectroscopy}

\subsection{Cyclotron null}

As we can see directly from Eq.~(\ref{eq:QuantumVRI}), the optical invariant is only nontrivial $|N|=2$ when the viscous cyclotron frequency $\Omega_c(q)=0$ changes sign [Eq.~(\ref{eq:VCF})]. We define this as the cyclotron null which occurs at the critical in-plane momentum $q=D_H^{-1}$ and cannot be removed unless there is a topological phase transition that closes the spectral gap. $q=D_H^{-1}$ is the momentum where the magnetic field is completely expelled from the fluid $B_\mathrm{eff}(q)=0$ and is unique to the nontrivial regime $\omega_c\nu_H>0$. At this particular momentum, the cyclotron motion switches handedness and the circulating currents appear to rotate in the opposite direction. This interesting phenomenon is reminiscent of the Meissner effect in a superconductor \cite{Hirsch_2012}, that causes all magnetic fields to be expelled from the electron fluid. The difference is that the cyclotron null $\Omega_c(q)=0$ is a deep sub-wavelength effect as it is a consequence of Hall viscosity. Moreover, since the gyrotropy $g\propto\Omega_c$ is directly proportional to $\Omega_c$, any null in $\Omega_c=0$ will manifest as a vanishing $g=0$. We identify this intriguing phenomenon as the experimental signature of nontrivial optical topology.

\begin{figure}
    \centering
    \includegraphics[width=0.9\linewidth]{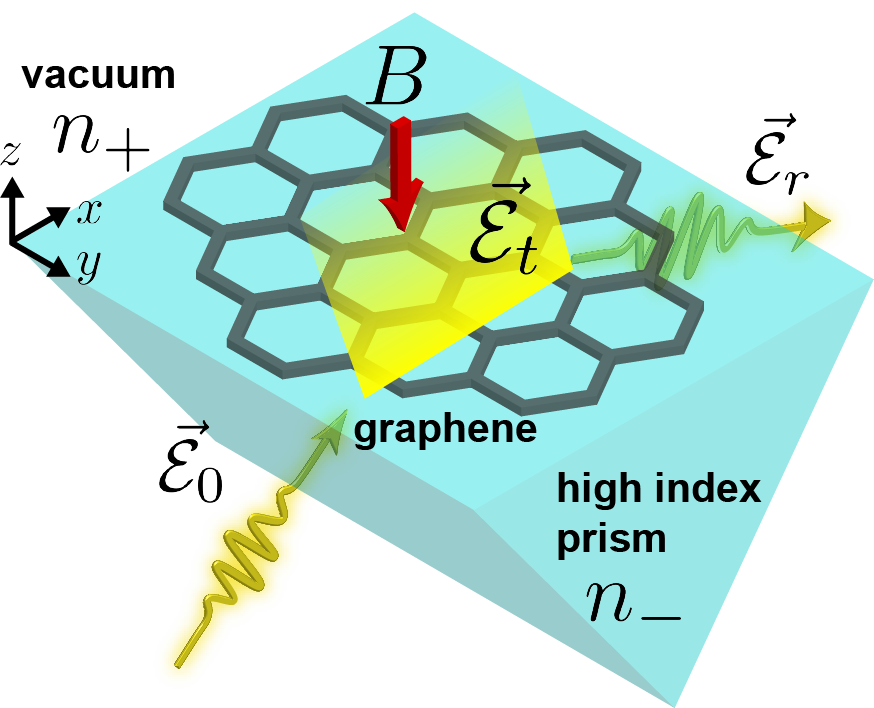}
    \caption{Experimental setup to probe e-MOKE of bulk monolayer graphene under static $B$-field. $n_+$ and $n_-$ are the refractive indices of the vacuum and prism respectively. A high index prism will be required to access the deep sub-wavelength gyrotropic effects as free space photons do not possess sufficient momentum.}
    \label{fig:Graphene}
\end{figure}

\subsection{Sub-wavelength $B$-field repulsion}

Our goal is to measure this magnetic field repulsion in the bulk material which is revealed in the reflection and transmission spectra of high-momentum photons. As the viscous Hall fluid manifests sub-wavelength cyclotron nulls $\Omega_c(q)=0$, only evanescent electromagnetic waves can probe their existence. A schematic of the evanescent magneto-optic Kerr effect (e-MOKE) spectroscopic probe is displayed in Fig.~\ref{fig:Graphene}. The configuration is graphene with an applied $B$-field on a dielectric substrate and the top exposed to vacuum. For incident light, we exploit a high index prism available at THz and lower frequencies to interface total internally reflected evanescent waves with the viscous Hall fluid. Although the local interaction at the sample location has large momentum, the sub-wavelength gyrotropic information is carried to the far field by reflected photons facilitating the read-out of the e-MOKE angle through traditional lock-in techniques.

\begin{figure}
    \centering
    \includegraphics[width=\linewidth]{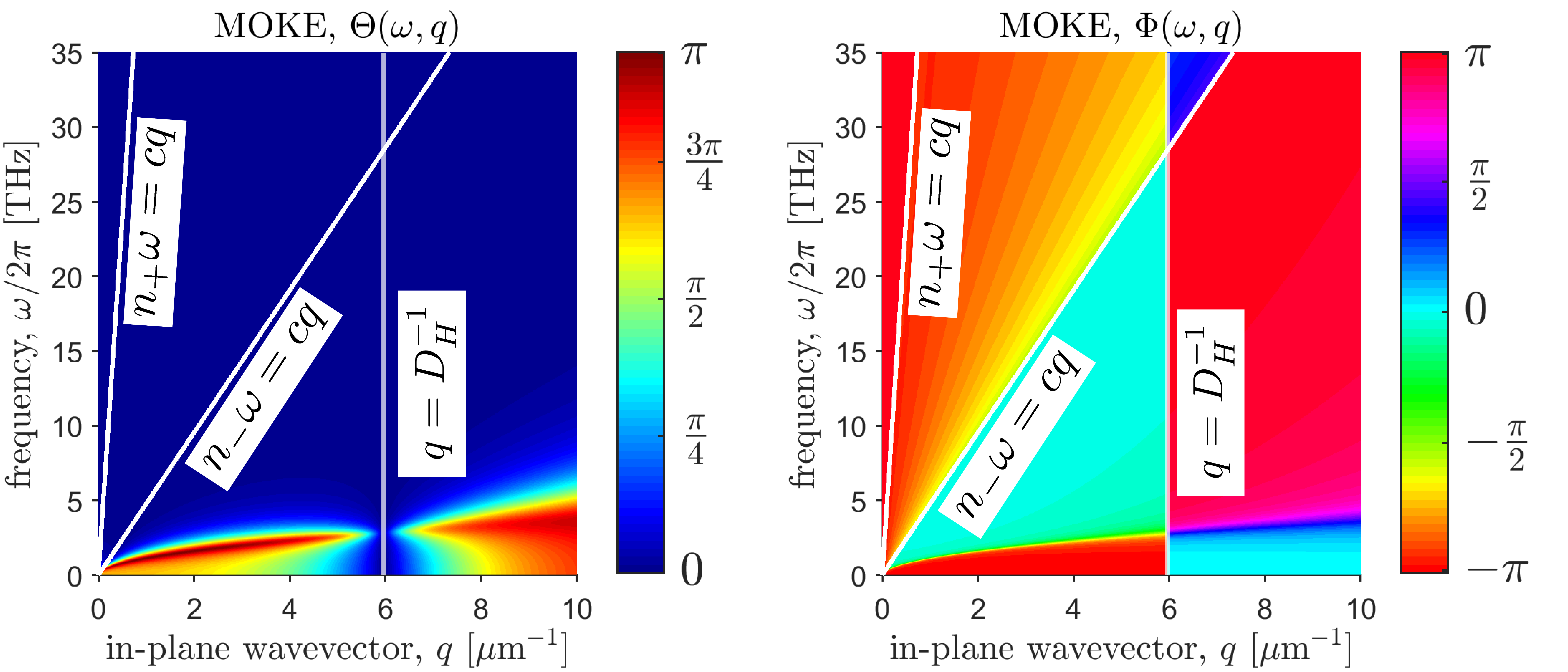}
    \caption{Probing the e-MOKE of the viscous Hall fluid by measuring the rotation of the reflected wave $\tan(\Theta/2) \exp(i\Phi)=r_{ps}/r_{ss}$. The bright peak is the surface plasmon. In the nontrivial (repulsive) regime $\omega_c\nu_H>0$, the MOKE vanishes $\Theta(q)=0$ at the critical momentum $q=D_H^{-1}$, where $D_H=\sqrt{\nu_H/\omega_c}$ is the Hall diffusion length. Traversing this cyclotron null causes the sense of rotation to reverse.}
    \label{fig:reflection}
\end{figure}

We explicitly calculate the relevant parameters for monolayer graphene under conventional laboratory settings to isolate the frequency-momentum space that should be explored for topological phenomena [Tbl.~\ref{tab:graphene}]. An incident photon $\vec{\mathcal{E}}_0$ will be reflected $\vec{\mathcal{E}}_r$ with a relationship between $\hat{s}$ and $\hat{p}$ polarizations,
\begin{equation}\label{eq:ReflectMatrix}
\begin{bmatrix}
\mathcal{E}_{rs}\\ \mathcal{E}_{rp}
\end{bmatrix}=\begin{bmatrix}
r_{ss}& r_{sp}\\
r_{ps} & r_{pp}
\end{bmatrix}\begin{bmatrix}
\mathcal{E}_{0s}\\ \mathcal{E}_{0p}
\end{bmatrix}.
\end{equation}
The exact expressions for the reflection coefficients are derived in the supplementary information along with a short review of boundary conditions on 2D charge densities. Importantly, the off-diagonal components, $r_{ps}$ and $r_{sp}$, couple $\hat{s}$ and $\hat{p}$ polarizations which means an incident linearly polarized wave will be elliptical upon reflection. This is known as the MOKE and it is directly proportional to the gyrotropy $r_{ps}=r_{sp}\propto g\propto \Omega_c$ \cite{Hamrle_2010}. Hence, to determine if the sample is topologically nontrivial, we must look for nulls in the MOKE rotation,
\begin{equation}
\tan\left(\frac{\Theta}{2}\right) \exp(i\Phi)=\frac{r_{ps}}{r_{ss}},
\end{equation}
which will occur at the critical in-plane momentum $\Theta(q)=0$. The MOKE rotation ($s$-effect) is defined by the ratio of reflected $\hat{p}$ to $\hat{s}$ polarization due to an incident $\hat{s}$ polarized wave. $\Theta\in[0,\pi]$ is the relative magnitude and $\Phi\in[-\pi,\pi]$ is the relative phase. 
Theoretical plots of the $s$-effect are displayed in Fig.~\ref{fig:reflection}.

\section{Conclusions}

We have introduced the optical phases $N\in\mathbb{Z}$ of two-dimensional quantum matter -- a topological classification emerging from the invariant optical proprieties of a material. As a particularly important example, we have shown that $N\neq 0$ is nontrivial in graphene's viscous Hall fluid and is fundamentally tied to the Hall viscosity $\eta_H$. We have also proposed a new probe for 2D materials: evanescent magneto-optical Kerr effect (e-MOKE) spectroscopy to search for this nontrivial phase of matter and the unique $B$-field expulsion that is reminiscent of the Meissner effect. These intriguing optical $N$-phases are also expected in viscous Chern insulators and viscous Weyl-semimetals, leading to a new generation of effects at the interface of topological photonics and condensed matter physics. 

\section*{Acknowledgements}

This research was supported by the Defense Advanced Research Projects Agency (DARPA) Nascent Light-Matter Interactions (NLM) Program. This material is based upon work partially supported by the U.S. Department of Energy, Office of Basic Energy Sciences [Grant. No. DE-SC0017717].

\bibliography{top_optics.bib}

\end{document}